\documentclass{ptephy_v1}

\preprintnumber{XXXX-XXXX} 

\usepackage{bm}
\usepackage{indentfirst}

\begin{document}

\title{Non-Bloch band theory and bulk-edge correspondence in non-Hermitian systems}


\author[1]{Kazuki Yokomizo}
\author[1,2]{Shuichi Murakami}
\affil[1]{Department of Physics, Tokyo Institute of Technology, 2-12-1 Ookayama, Meguro-ku, Tokyo, 152-8551, Japan\email{k-yoko@stat.phys.titech.ac.jp}}
\affil[2]{TIES, Tokyo Institute of Technology, 2-12-1 Ookayama, Meguro-ku, Tokyo, 152-8551, Japan} 


\begin{abstract}%
In this paper, we review our non-Bloch band theory in one-dimensional non-Hermitian tight-binding systems. In our theory, it is shown that in non-Hermitian systems, the Brillouin zone is determined so as to reproduce continuum energy bands in a large open chain. By using simple models, we explain the concept of the non-Bloch band theory and the method to calculate the Brillouin zone. In particular, for the non-Hermitian Su-Schrieffer-Heeger model, the bulk-edge correspondence can be established between the topological invariant defined from our theory and existence of the topological edge states.
\end{abstract}

\subjectindex{xxxx, xxx}

\maketitle

%
%

\section{\label{sec1}Introduction}
In recent years, interest in studies of non-Hermitian systems has been rapidly growing both in theories and in experiments. The non-Hermitian Hamiltonian is useful for studying non-equilibrium systems and open systems, which exchange energies and particles with external environment. Non-Hermitian systems emerge in various fields of classical physics and quantum physics~\cite{1}. In classical systems, gain and loss lead to non-Hermitian terms in eigenvalue equations. On the other hand, in quantum systems, one of the origins of non-Hermiticity is many-body correlation effect. For example, in strongly correlated electron systems, we can get non-Hermitian Hamiltonian by incorporating the imaginary part of the self-energy representing the lifetime of quasiparticle into one-body Hamiltonian~\cite{2,3,4}. One of the most intriguing topics is how non-Hermitian effects affect topological physics. For example, some previous works proposed new definitions of a gap in non-Hermitian systems, and in terms of this definition, topological classifications of gapped phases and gapless phases are given under some symmetries~\cite{4,5,6,7,8,9,10,11}. Furthermore, new topological invariants can appear thanks to unique features of non-Hermitian systems such as non-Hermitian degeneracies~\cite{12}.

Among theoretical works of non-Hermitian topological systems, in particular, violation of the bulk-edge correspondence has been a long-standing issue in this field, and the reasons for this violation have been under debate~\cite{13,14,15,16,17,18,19}. One of the controversies is that in most of the previous works, the Bloch wave number $k$ has been treated as real in non-Hermitian systems, similarly to Hermitian ones. In contrast to these previous works, it was proposed that $k$ takes complex values in non-Hermitian systems in order to describe electronic states in a long open chain because the energy spectrum in a periodic chain and that in an open chain are different~\cite{20,21}. Then the value of $\beta\equiv{\rm e}^{ik}$ is confined on a loop on the complex plane so as to reproduce continuum energy bands in a large open chain. This loop is a generalization of the conventional Brillouin zone and is called generalized Brillouin zone (GBZ). Furthermore the eigenstates of the non-Hermitian Hamiltonian do not necessarily extend over the whole system but are localized at either end of the chain. This phenomenon is called the non-Hermitian skin effect~\cite{20}. After the proposal of this effect, intriguing properties of the non-Hermitian skin effect have been studied intensively both in theories and experiments~\cite{22,23,24,25,26,27,28,29}.

In our work~\cite{21}, we established a non-Bloch band theory in a one-dimensional (1D) non-Hermitian tight-binding system. We showed how to determine the GBZ $C_\beta$ for $\beta\equiv{\rm e}^{ik},~k\in{\mathbb C}$. In the present paper, we review this non-Bloch band theory with simple examples. In order to understand the concept of our theory, we explain in detail how $C_\beta$ can be obtained in simple models. Furthermore we can establish the bulk-edge correspondence in the non-Hermitian Su-Schrieffer-Heeger (SSH) model from the topological invariant defined by $C_\beta$.

%
%

\section{\label{sec2}Simple model}
\begin{figure}[!h]
\centering
\includegraphics[width=15.3cm]{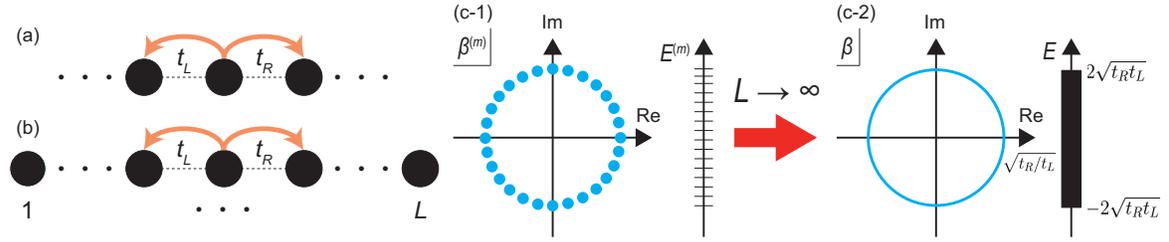}
\caption{\label{fig1}(a) Simple model in an infinite open chain. (b) Simple model in a finite open chain with the system size $L$. (c-1) Schematic figures of the distribution of $\beta^{\left(m\right)}$ and that of $E^{\left(m\right)}$ with $m=1,\cdots,L$. In the limit of $L\rightarrow\infty$, $\beta^{\left(m\right)}$ forms a circle with the radius $\sqrt{t_R/t_L}$, and $E^{\left(m\right)}$ forms the continuum energy band with the range of $\left[-2\sqrt{t_Rt_L},2\sqrt{t_Rt_L}\right]$, shown in (c-2).}
\end{figure}
In this section, we study the constructions of the GBZ and of the continuum energy bands in a simple non-Hermitian tight-binding model. This model is known as the Hatano-Nelson model without disorder~\cite{30}. The real-space Hamiltonian of this system is given by
\begin{equation}
H=\sum_{n=1}^{L-1}\left(t_Rc_{n+1}^\dag c_n+t_Lc_n^\dag c_{n+1}\right),
\label{eqSim1}
\end{equation}
where $t_R,t_L\in{\mathbb R}$ are nearest-neighbor asymmetric hopping amplitudes to the right and to the left, respectively (Fig.~\ref{fig1}(a)). We show the schematic figure of this system in a finite open chain with the system size $L$ in Fig.~\ref{fig1}(b). The real-space eigen-equation $H|\psi\rangle=E|\psi\rangle$, for the eigenvector $|\psi\rangle=\left(\psi_1,\cdots,\psi_L\right)^{\rm T}$, can be written as
\begin{equation}
t_R\psi_{n-1}+t_L\psi_{n+1}=E\psi_n,~(n=1,\cdots,L),~\psi_0=\psi_{L+1}=0.
\label{eqSim2}
\end{equation}
From the theory of linear difference equations, a general solution of the recursion equation (\ref{eqSim2}) is written as
\begin{equation}
\psi_n=\left(\beta_1\right)^n\phi^{\left(1\right)}+\left(\beta_2\right)^n\phi^{\left(2\right)},
\label{eqSim3}
\end{equation}
where $\beta_j~(j=1,2)$ are the solutions of the equation
\begin{equation}
t_R\beta^{-1}+t_L\beta=E.
\label{eqSim4}
\end{equation}
Together with the open boundary conditions $\psi_0=0$ and $\psi_{L+1}=0$, we can get
\begin{equation}
\left( \begin{array}{cc}
1                          & 1                          \\
\left(\beta_1\right)^{L+1} & \left(\beta_2\right)^{L+1}
\end{array}\right)\left( \begin{array}{c}
\phi^{\left(1\right)} \\
\phi^{\left(2\right)}
\end{array}\right)=\left( \begin{array}{c}
0 \\
0
\end{array}\right)
\label{eqSim5}
\end{equation}
and obtain a boundary equation, which represents the boundary conditions at the two ends, as $\left(\beta_1/\beta_2\right)^{L+1}=1$ so that the coefficients $\phi^{\left(1\right)},\phi^{\left(2\right)}$ take nonzero values. Then one can get 
\begin{equation}
\frac{\beta_1}{\beta_2}={\rm e}^{2i\theta_m},~\left(\theta_m=\frac{m\pi}{L+1},~m=1,\cdots,L\right).
\label{eqSim6}
\end{equation}
Therefore, from Eq.~(\ref{eqSim4}), $\beta_j~(j=1,2)$ can be written as
\begin{equation}
\beta_1^{\left(m\right)}=r{\rm e}^{i\theta_m},~\beta_2^{\left(m\right)}=r{\rm e}^{-i\theta_m},
\label{eqSim7}
\end{equation}
where $r=\left|\beta_{1,2}\right|=\sqrt{t_R/t_L}$ because $\beta_1\beta_2=t_R/t_L$ from the Vieta's formula, and the eigenstate (\ref{eqSim3}) and the eigenenergy (\ref{eqSim4}) are written as
\begin{equation}
\psi_n^{(m)}\propto\left(r{\rm e}^{i\theta_m}\right)^n-\left(r{\rm e}^{-i\theta_m}\right)^n\propto r^n\sin n\theta_m,~E^{\left(m\right)}=2\sqrt{t_Rt_L}\cos\theta_m,
\label{eqSim8}
\end{equation}
respectively. From Eq.~(\ref{eqSim8}), the distribution of the discrete eigenstates and that of the discrete energy levels are shown in Fig.~\ref{fig1}(c-1).

Now, as the system size $L$ becomes larger, these eigenstates and energy levels become dense. Finally, in the limit of $L\rightarrow\infty$ (Fig.~\ref{fig1}(a)), the energy levels form a continuum energy band as shown in Fig.~\ref{fig1}(c-2), leading to the form of the continuum eigenstates as
\begin{equation}
\beta_1=\sqrt{\frac{t_R}{t_L}}{\rm e}^{i\theta},~\beta_2=\sqrt{\frac{t_R}{t_L}}{\rm e}^{-i\theta}
\label{eqSim9}
\end{equation}
from Eq.~(\ref{eqSim7}), and to that of the continuum energy band as
\begin{equation}
E=2\sqrt{t_Rt_L}\cos\theta
\label{eqSim10}
\end{equation}
from Eq.~(\ref{eqSim4}) by changing the parameter $\theta\in{\mathbb R}$.

By a comparison with a Hermitian case, which is realized when $t_R=t_L$, we can intuitively understand these results in the viewpoint of the Bloch band theory. Equation (\ref{eqSim3}) means that $\beta(=\beta_1,\beta_2)$ can be related with the Bloch wave number $k$ by $\beta={\rm e}^{ik}$. In this sense, the distribution of $\beta(={\rm e}^{ik})$, which is called generalized Brillouin zone (GBZ), gives a non-Hermitian extension of the Brillouin zone. In 1D Hermitian systems, the wave number is real, and the GBZ is always a unit circle. On the other hand, in the present case, the Brillouin zone is not a unit circle, meaning that the corresponding wave number is not real. We note that the eigenstate for the finite chain of the simple model is a superposition of two ``plane waves'' with $\beta_1$ and $\beta_2$, and these two values satisfy Eq.~(\ref{eqSim4}) with the same energy $E$. Therefore, once the GBZ is shown, the energy eigenvalues $E$ are calculated from Eq.~(\ref{eqSim4}). As a result, the continuum energy band is formed in the range of $\left[-2\sqrt{t_Rt_L},2\sqrt{t_Rt_L}\right]$.

Here we comment the dependence of the above results on boundary conditions. If we change the boundary conditions, $\psi_0=a$ and $\psi_{L+1}=b$ as an example, the form of the boundary equation in a finite open chain is modified. Then the energy levels (\ref{eqSim8}) are also modified, which means that the energy levels in a finite open chain depend on boundary conditions. Nevertheless, in the limit of $L\rightarrow\infty$, the continuum energy bands and the GBZ are independent of boundary conditions in an open chain. Therefore they can be obtained as shown in Fig.~\ref{fig1}(c-2) under any boundary conditions in an open chain.

%
%

\section{\label{sec3}Non-Bloch band theory}

%
%

\subsection{\label{sec3-1}Concept}
\begin{figure}[!h]
\centering
\includegraphics[width=15.3cm]{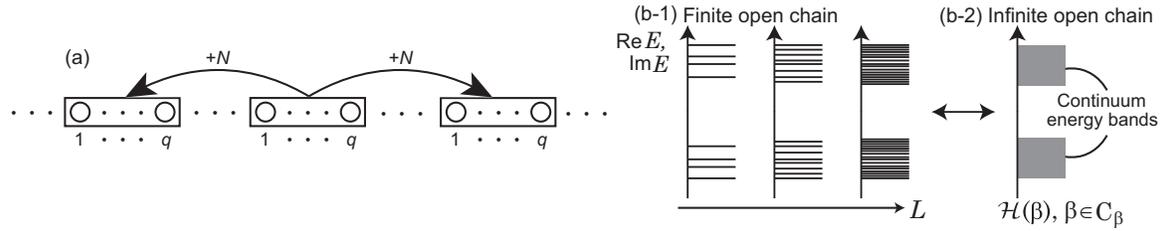}
\caption{\label{fig2}(a) One-dimensional tight-binding system. A unit cell includes $q$ degree of freedom, and the range of hopping is $N$. (b) Schematic figures of (b-1) energy levels in a finite open chain with various system sizes $L$ and of (b-2) continuum energy bands in an infinite open chain. The vertical axis represents the distribution of the complex energy $E$.}
\end{figure}
We can generalize the result in the previous section to general 1D non-Hermitian systems. We start with a 1D tight-binding system with spatial periodicity as shown in Fig.~\ref{fig2}(a). A unit cell is composed of $q$ degrees of freedom, such as sublattices, spins, or orbitals, and the electrons hop to the $N$-th nearest unit cells. Then its Hamiltonian can be written as
\begin{equation}
H=\sum_n\sum_{i=-N}^N\sum_{\mu,\nu=1}^qt_{i,\mu\nu}c_{n+i,\mu}^\dag c_{n,\nu},
\label{eqNon1}
\end{equation}
where $c_{n,\mu}^\dag~(c_{n,\mu})$ is a creation (an annihilation) operator of an electron with an index $\mu~(\mu=1,\cdots,q)$ in the $n$-th unit cell, and $t_{i,\mu\nu}$ is a hopping amplitude to the $i$-th nearest unit cell. This Hamiltonian can be non-Hermitian, meaning that $t_{i,\mu\nu}$ is not necessarily equal to $t_{-i,\nu\mu}^\ast$. In this situation, the real-space eigen-equation is written as $H|\psi\rangle=E|\psi\rangle$, where the eigenvector is given by $|\psi\rangle=(\cdots,\psi_{1,1},\cdots,\psi_{1,q},\psi_{2,1},\cdots,\psi_{2,q},\cdots)^{\rm T}$. Then, as is similar to the model in the previous section, $|\psi\rangle$ can be represented as a linear combination:
\begin{equation}
\psi_{n,\mu}=\sum_j\phi_{n,\mu}^{\left(j\right)},~\phi_{n,\mu}^{\left(j\right)}=\left(\beta_j\right)^n\phi_\mu^{\left(j\right)},~\left(\mu=1,\cdots,q\right),
\label{eqNon2}
\end{equation}
where $\beta=\beta_j$ is a solution of the characteristic equation defined as
\begin{equation}
\det\left[{\cal H}\left(\beta\right)-E\right]=0,~\left[{\cal H}\left(\beta\right)\right]_{\mu\nu}=\sum_{i=-N}^Nt_{i,\mu\nu}\left(\beta\right)^i,~\left(\mu,\nu=1,\cdots,q\right),
\label{eqNon3}
\end{equation}
and the eigen-equation of the matrix ${\cal H}\left(\beta\right)$ can be explicitly rewritten as
\begin{equation}
\sum_{\nu=1}^q\left[{\cal H}\left(\beta\right)\right]_{\mu\nu}\phi_\nu=E\phi_\nu,~\left(\mu=1,\cdots,q\right).
\label{eqNon4}
\end{equation}
We note that ${\cal H}\left(\beta\right)$ becomes the Bloch Hamiltonian if we rewrite it in terms of the conventional Bloch wave number $k$. Furthermore the equation (\ref{eqNon3}) is an algebraic equation for $\beta$ with an even degree $2M=2qN$.

Now we explain the concept of the non-Bloch band theory. The resulting energy levels are discrete in a finite open chain with a system size $L$ as shown in Fig.~\ref{fig2}(b-1). Here, as $L$ becomes larger, the energy levels become dense and asymptotically continuous. Finally, in the limit of $L\rightarrow\infty$, the continuum energy bands are formed as shown in Fig.~\ref{fig2}(b-2). Then the asymptotic distribution of $\beta$ for $L\rightarrow\infty$ is the GBZ $C_\beta$. We note that in this case, the absolute value of $\beta$ is not necessarily unity, and $C_\beta$ is obtained as a loop on the complex plane. The key question is how to construct the GBZ for the system considered here. From the argument so far, we may need to calculate the energy levels for finite $L$ and study its asymptotic behavior for $L\rightarrow\infty$. This is a cumbersome procedure, and the result may possibly depend on boundary conditions.

Remarkably, in Ref.~\cite{21}, we found a method to calculate the GBZ $C_\beta$, without going through a calculation on a finite open chain with the system size $L$. This largely simplifies the calculation. It is worth noting that $C_\beta$ is independent of boundary conditions. Thus, while energy levels in a finite open chain depend on boundary conditions, their asymptotic behaviors do not.

Below we explain a way to calculate the GBZ $C_\beta$, which determines continuum energy bands. Let $\beta_j~(j=1,\cdots,2M)$ be the solutions of the equation (\ref{eqNon3}). When we number the $2M$ solutions so as to satisfy
\begin{equation}
\left|\beta_1\right|\leq\left|\beta_2\right|\leq\cdots\leq\left|\beta_{2M-1}\right|\leq\left|\beta_{2M}\right|,
\label{eqNon5}
\end{equation}
we find that the condition for continuum energy bands is given by
\begin{equation}
\left|\beta_M\right|=\left|\beta_{M+1}\right|,
\label{eqNon6}
\end{equation}
and the trajectories of $\beta_M$ and $\beta_{M+1}$ give $C_\beta$. The example in Sec.~\ref{sec2} is a special case with $M=1$. In Secs.~\ref{sec3-2} and \ref{sec3-3}, we will show some examples of $C_\beta$ and the continuum energy bands calculated by using the condition (\ref{eqNon6}). Here we note that although the eigenenergies for the continuum energy bands are obtained from Eq.~(\ref{eqNon3}) by putting $\beta=\beta_M$ and $\beta=\beta_{M+1}$, the eigenvectors of Eq.~(\ref{eqNon4}) are not eigenstates of the Hamiltonian (\ref{eqNon1}). Instead, the eigenstates of the Hamiltonian (\ref{eqNon1}) is given by Eq.~(\ref{eqNon2}), which involves the terms with $\beta=\beta_1,\cdots,\beta_{2M}$. Nonetheless, the non-Bloch band theory explained here says that the eigenenergies for the continuum energy bands are determined by $\beta_M$ and $\beta_{M+1}$, and that the GBZ $C_\beta$ and a set of the eigenenergies are independent of boundary conditions in an open chain. Thus, in the calculation of $C_\beta$, we do not need to solve the eigenvalue problem in Eq.~(\ref{eqNon4}). The matrix ${\cal H}\left(\beta\right)$ is introduced here in order to express the characteristic equation (\ref{eqNon3}).

While the conclusion in Eq.~(\ref{eqNon6}) is simple, its derivation presented in the Supplemental Material of Ref.~\cite{21} is lengthy. Therefore, instead of reproducing it here, we explain its outline. First, we impose the given boundary conditions in an open chain onto the eigenvector $|\psi\rangle$ with Eq.~(\ref{eqNon2}) to get a set of the linear equations for $\phi_\mu^{\left(j\right)}$. Then the condition for this set of the linear equations to have a nontrivial solution yields an equation for $\beta_j$'s, called boundary equation. This boundary equation is a complicated equation dependent on the boundary conditions, giving discrete energy levels. Nonetheless, in the limit of a large system size, $L\rightarrow\infty$, we expect that the energy levels become dense and eventually form continuum energy bands. Therefore we impose a condition that the solutions of the boundary equation should have an asymptotically dense set. Then we get the Eq.~(\ref{eqNon6}) which is eventually independent of the form of open boundary conditions. More details are presented in the Supplemental Material of Ref.~\cite{21}.

The condition for continuum energy bands (\ref{eqNon6}) can be regarded as a condition for formation of a standing wave. Equation~(\ref{eqNon6}) means that the decay lengths of the eigenstates corresponding to $\beta_M$ and $\beta_{M+1}$ are equal, so that the wave function vanishes at both ends of an open chain. For example, in the model in Sec.~\ref{sec2}, the wave function (\ref{eqSim8}) represents a standing wave apart from the factor $r^n$, as a superposition of two counterpropagating ``plane waves''. Furthermore the condition (\ref{eqNon6}) is physically reasonable in several aspects. Firstly, this condition does not depend on any boundary conditions in an open chain. Secondly, in the Hermitian limit, we can rewrite Eq.~(\ref{eqNon6}) to the well known result, i.e. $\left|\beta_M\right|=\left|\beta_{M+1}\right|=1$. For example, in the model in Sec.~\ref{sec2} with the case of $t_R=t_L$, the model becomes Hermitian, and the GBZ becomes a unit circle, identified with the conventional Brillouin zone.

Finally we mention the case that the characteristic equation (\ref{eqNon3}) is a reducible algebraic equation. Namely it can be factorized as $\det\left[{\cal H}\left(\beta\right)-E\right]=f_1\left(\beta,E\right)\cdots f_m\left(\beta,E\right)$, where $f_i\left(\beta,E\right)~(i=1,\cdots,m)$ are algebraic equations for $\beta$ and $E$. For simplicity, we assume that they are algebraic equations for $\beta$ with an even degree $2M_i$. In this case, the continuum energy bands and the GBZs can be obtained from the conditions $\left|\beta_{M_i}\right|=\left|\beta_{M_i+1}\right|~(i=1,\cdots,m)$ instead of Eq.~(\ref{eqNon6})~\cite{31}.

%
%

\subsection{\label{sec3-2}Non-Hermitian SSH model}
\begin{figure}[!h]
\centering
\includegraphics[width=15.3cm]{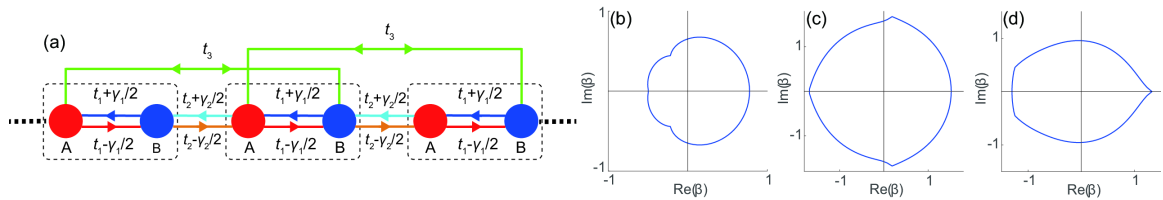}
\caption{\label{fig3}Non-Hermitian Su-Schrieffer-Heeger (SSH) model in an infinite open chain. The dotted boxes indicate the unit cell. (b)-(d) Generalized Brillouin zone in the non-Hermitian SSH model. The values of the parameters are (b) $t_1=1.1,t_2=1,t_3=1/5,\gamma_1=4/3$, and $\gamma_2=0$; (c) $t_1=0.3,t_2=1.1,t_3=1/5,\gamma_1=0$, and $\gamma_2=-4/3$; and (d) $t_1=-0.3,t_2=0.5,t_3=1/5,\gamma_1=5/3$, and $\gamma_2=1/3$.}
\end{figure}
In this subsection, in terms of the non-Bloch band theory, we investigate the non-Hermitian SSH model, which has been studied in some previous works~\cite{16,19,20,21,28}. The schematic figure of the non-Hermitian SSH model is shown in Fig.~\ref{fig3}(a), and the real-space Hamiltonian is written as
\begin{eqnarray}
H&=&\sum_n\left[t_3c_{n,{\rm A}}^\dag c_{n+1,{\rm B}}+\left(t_1+\frac{\gamma_1}{2}\right)c_{n,{\rm A}}^\dag c_{n,{\rm B}}+\left(t_2-\frac{\gamma_2}{2}\right)c_{n+1,{\rm A}}^\dag c_{n,{\rm B}}\right. \nonumber\\
&&+\left.\left(t_2+\frac{\gamma_2}{2}\right)c_{n,{\rm B}}^\dag c_{n+1,{\rm A}}+\left(t_1-\frac{\gamma_1}{2}\right)c_{n,{\rm B}}^\dag c_{n,{\rm A}}+t_3c_{n+1,{\rm B}}^\dag c_{n,{\rm A}}\right].
\label{eqNon7}
\end{eqnarray}
Henceforth we set all the parameters to be real. For the real-space eigenvector $|\psi\rangle=\left(\cdots,\psi_{1,{\rm A}},\psi_{1,{\rm B}},\psi_{2,{\rm A}},\psi_{2,{\rm B}},\cdots\right)$, we can explicitly write the real-space eigen-equation as
\begin{eqnarray}
\left\{ \begin{array}{l}
\displaystyle t_3\psi_{n+1,{\rm B}}+\left(t_1+\frac{\gamma_1}{2}\right)\psi_{n,{\rm B}}+\left(t_2-\frac{\gamma_2}{2}\right)\psi_{n-1,{\rm B}}=E\psi_{n,{\rm A}}, \vspace{8pt}\\
\displaystyle \left(t_2+\frac{\gamma_2}{2}\right)\psi_{n+1,{\rm A}}+\left(t_1-\frac{\gamma_1}{2}\right)\psi_{n,{\rm A}}+t_3\psi_{n-1,{\rm A}}=E\psi_{n,{\rm B}}.
\end{array}\right.
\label{eqNon8}
\end{eqnarray}
Here we can take a general ansatz for the wave function as a linear combination:
\begin{equation}
\left( \begin{array}{c}
\psi_{n,{\rm A}} \\
\psi_{n,{\rm B}}
\end{array}\right)=\sum_j\left(\beta_j\right)^n\left( \begin{array}{c}
\phi_{\rm A}^{\left(j\right)} \\
\phi_{\rm B}^{\left(j\right)}
\end{array}\right),
\label{eqNon9}
\end{equation}
where $\beta=\beta_j$ are the solutions of the characteristic equation $\det\left[{\cal H}\left(\beta\right)-E\right]=0$ for the matrix
\begin{eqnarray}
{\cal H}\left(\beta\right)=\left( \begin{array}{cc}
0                                                                                                        & \displaystyle t_3\beta+\left(t_1+\frac{\gamma_1}{2}\right)+\left(t_2-\frac{\gamma_2}{2}\right)\beta^{-1} \\
\displaystyle \left(t_2+\frac{\gamma_2}{2}\right)\beta+\left(t_1-\frac{\gamma_1}{2}\right)+t_3\beta^{-1} & 0
\end{array}\right).
\label{eqNon10}
\end{eqnarray}
In this case, the characteristic equation
\begin{equation}
\left[t_3\beta+\left(t_1+\frac{\gamma_1}{2}\right)+\left(t_2-\frac{\gamma_2}{2}\right)\beta^{-1}\right]\left[\left(t_2+\frac{\gamma_2}{2}\right)\beta+\left(t_1-\frac{\gamma_1}{2}\right)+t_3\beta^{-1}\right]-E^2=0
\label{eqNon11}
\end{equation}
is a quartic equation for $\beta$, having four solutions $\beta_i~(i=1,\cdots,4)$ satisfying $\left|\beta_1\right|\leq\left|\beta_2\right|\leq\left|\beta_3\right|\leq\left|\beta_4\right|$. In this case of $M=2$, the condition for continuum energy bands is given by
\begin{equation}
\left|\beta_2\right|=\left|\beta_3\right|.
\label{eqNon12}
\end{equation}
It is obtained by imposing that Eq.~(\ref{eqNon9}) satisfying open boundary conditions forms a dense set of solutions at $L\rightarrow\infty$. As emphasized earlier, Eq.~(\ref{eqNon12}) does not depend on boundary conditions in an open chain as shown in the Supplemental Material of Ref.~\cite{21}.

Here the trajectories of $\beta_2$ and $\beta_3$ give the GBZ $C_\beta$ as shown in Figs.~\ref{fig3}~(b)-(d) with various values of the parameters. It is worth mentioning some features of $C_\beta$ in the following. First, as shown in Fig.~\ref{fig3}(d), $\left|\beta\right|$ on $C_\beta$ takes both values more than $1$ and values less than $1$~\cite{23,24,31}. Here, $\left|\beta\right|>1~\left(\left|\beta\right|<1\right)$ means that the eigenstate is localized at the right (left) end of the chain, representing the non-Hermitian skin effect. Second, $C_\beta$ is a closed loop encircling the origin on the complex plane~\cite{24,31}. Finally, $C_\beta$ can have the cusps, corresponding to the cases where three solutions of Eq.~(\ref{eqNon11}) share the same absolute value.

In order to calculate the GBZ, we need to judge the condition (\ref{eqNon12}) for a given value of $E$. To explain this, we set the values of the parameters as $\left(t_1,t_2,t_3,\gamma_1,\gamma_2\right)=\left(3/10,1/2,1/5,5/3,1/3\right)$. For example, when we take $E=E_1=0.140+0.755i$, we obtain the four solutions of Eq.~(\ref{eqNon11}) as
\begin{eqnarray}
\begin{array}{l}
\left(\beta_1,\beta_2,\beta_3,\beta_4\right)=\left(0.142-0.348i,0.064+0.535i,-0.506+0.184i,-4.57-0.371i\right),                       \vspace{5pt}\\
\left(\left|\beta_1\right|,\left|\beta_2\right|,\left|\beta_3\right|,\left|\beta_4\right|\right)=\left(0.376,0.538,0.538,4.58\right).
\end{array}
\label{eqNon13}
\end{eqnarray}
Since these solutions satisfy the condition for continuum energy bands (\ref{eqNon12}), we conclude that $E_1$ is included in the continuum energy bands. On the other hand, when we substitute $E=E_2=0.180+1.23i$ into Eq.~(\ref{eqNon11}), we obtain the solutions as
\begin{eqnarray}
\begin{array}{l}
\left(\beta_1,\beta_2,\beta_3,\beta_4\right)=\left(0.036-0.228i,-0.052+0.233i,-1.52+1.85i,-3.33-1.85i\right),                        \vspace{5pt}\\
\left(\left|\beta_1\right|,\left|\beta_2\right|,\left|\beta_3\right|,\left|\beta_4\right|\right)=\left(0.230,0.238,2.39,3.81\right).
\end{array}
\label{eqNon14}
\end{eqnarray}
In this case, $E_2$ is not in the continuum energy bands because $\left|\beta_2\right|\neq\left|\beta_3\right|$.

From Eq.~(\ref{eqNon12}), we calculate the continuum energy bands as shown in Fig.~\ref{fig5}(c). In fact, we can confirm that this result agrees with the energy levels in a finite open chain as shown in Fig.~\ref{fig5}(d). In conclusion, it is shown that Eq.~(\ref{eqNon12}) is appropriate for the condition for continuum energy bands. In Sec.~\ref{sec4}, we investigate the topological edge states appearing in a finite open chain as shown in red in Fig.~\ref{fig5}(d).

%
%

\subsection{\label{sec3-3}Simple two-band model}
\begin{figure}[!h]
\centering
\includegraphics[width=15.3cm]{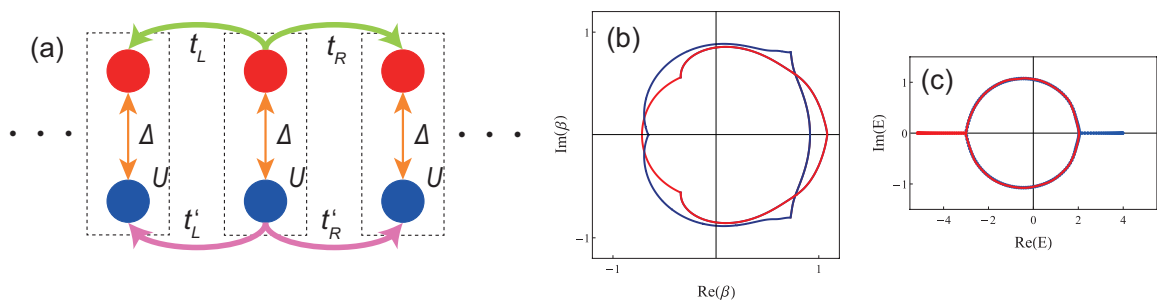}
\caption{\label{fig4}(a) The simple two-band model (\ref{eqNon15}) in an infinite open chain. The dotted boxes indicate the unit cell. (b) Generalized Brillouin zones and (c) continuum energy bands in the two-band model (\ref{eqNon15}). They correspond to each other in the same color. We set the values of the parameters as $t_R=2,t_L=1,t_R^\prime=1,t_L^\prime=3$, and $U=\Delta=-1$.}
\end{figure}
The previous work proposed that in non-Hermitian multi-band systems, each of the bands is associated with their own GBZ, from the condition for continuum energy bands (\ref{eqNon6})~\cite{31}. On the other hand, in Hermitian systems, all the bands have the same GBZ, being a unit circle $\beta={\rm e}^{ik},~k\in{\mathbb R}$. Here we show the splitting of the GBZ in a two-band non-Hermitian model as shown in Fig.~\ref{fig4}(a), following Ref.~\cite{31}. This system has no symmetries. The real-space Hamiltonian can be written as
\begin{eqnarray}
H&=&\sum_n\left(t_Rc_{n+1,{\rm A}}^\dag c_{n,{\rm A}}+\Delta c_{n,{\rm A}}^\dag c_{n,{\rm B}}+t_Lc_{n+1,{\rm A}}^\dag c_{n,{\rm A}}\right. \nonumber\\
&&+\left.t_R^\prime c_{n+1,{\rm B}}^\dag c_{n,{\rm B}}+\Delta c_{n,{\rm B}}^\dag c_{n,{\rm A}}+Uc_{n,{\rm B}}^\dag c_{n,{\rm B}}+t_L^\prime c_{n,{\rm B}}^\dag c_{n+1,{\rm B}}\right).
\label{eqNon15}
\end{eqnarray}
By the same procedure in Sec.~\ref{sec3-2}, we can get the matrix ${\cal H}\left(\beta\right)$ as
\begin{eqnarray}
{\cal H}\left(\beta\right)=\left( \begin{array}{cc}
t_R\beta^{-1}+t_L\beta & \Delta \\
\Delta                 & t_R^\prime\beta^{-1}+U+t_L^\prime\beta
\end{array}\right).
\label{eqNon16}
\end{eqnarray}
In this case, we can get the GBZs and the continuum bands by applying the condition for continuum energy bands (\ref{eqNon12}) to the solutions of the characteristic equation $\det\left[{\cal H}\left(\beta\right)-E\right]=0$, and the results are shown in Figs.~\ref{fig4}(b) and (c). One can see that the GBZs split into two curves, each of which corresponds to the individual band.  

We note that under some additional symmetries, some bands necessarily share the same GBZ. For example, the non-Hermitian SSH model (\ref{eqNon7}) has only one GBZ because it has sublattice symmetry (SLS). Here the SLS is defined as $\Gamma H\Gamma^{-1}=-H$ for a real-space Hamiltonian (\ref{eqNon1}), where $\Gamma$ is a unitary matrix satisfying $\Gamma^2=+1$. Namely the eigenenergy of this system appears in pairs, $\left(E,-E\right)$, both of which come from the same GBZ because of the form of the eigenvalue equation (\ref{eqNon11}).

%
%

\section{\label{sec4}Bulk-edge correspondence}
\begin{figure}[!h]
\centering
\includegraphics[width=15.3cm]{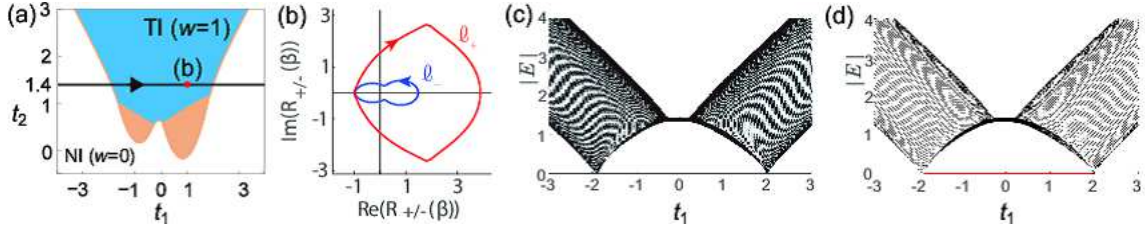}
\caption{\label{fig5}Bulk-edge correspondence in the non-Hermitian Su-Schrieffer-Heeger model with the values of the parameters as $t_3=1/5,\gamma_1=5/3$, and $\gamma_2=1/3$. (a) Phase diagram on the $t_1$-$t_2$ plane. The blue and the white regions represent a topological insulator phase with the winding number $w$ being $1$ and a normal insulator phase with $w=0$, respectively. The orange region represents the gapless phase. (b) Trajectories $\ell_+$ (red) and $\ell_-$ (blue) on the ${\bm R}$ plane with $t_1=1$ and $t_2=1.4$. The arrows mean the direction of the change of $R_\pm\left(\beta\right)$ as $\beta$ goes in a counterclockwise manner along the generalized Brillouin zone (GBZ). (c) Continuum energy bands from the GBZ. We show the results for them along the black arrow in (a) with $t_2=1.4$. (d) Energy levels in a finite open chain with the system size $L=100$. The red line represents the topological edge states.}
\end{figure}
In this section, we discuss the bulk-edge correspondence in the non-Hermitian SSH model introduced in Sec.~\ref{sec3-2}. It is shown that the topological invariant defined in terms of the GBZ can precisely predict existence of the topological edge states.

First of all, we define a topological invariant in a 1D non-Hermitian system with the SLS in a two-band model. Let us start with the real-space Hamiltonian (\ref{eqNon1}). By the procedure explained in Sec.~\ref{sec3-1}, we can get the matrix ${\cal H}\left(\beta\right)$ in the form of Eq.~(\ref{eqNon3}) in the bulk of a large open chain. Then, if we can put the matrix form of the SLS as ${\rm diag}(1,-1)$, we can rewrite ${\cal H}\left(\beta\right)$ of this system as an off-diagonal form;
\begin{eqnarray}
{\cal H}\left(\beta\right)=\left( \begin{array}{cc}
0                     & R_+\left(\beta\right) \\
R_-\left(\beta\right) & 0
\end{array}\right),
\label{eqBul1}
\end{eqnarray}
where $R_\pm\left(\beta\right)$ are polynomials of $\beta$ and $\beta^{-1}$. The complex wave number can be determined as $\beta\equiv{\rm e}^{ik},~k\in{\mathbb C}$ on the GBZ $C_\beta$ given from Eq.(\ref{eqNon6}). We note that the energy eigenvalues can be explicitly written as $E_\pm\left(\beta\right)=\pm\sqrt{R_+\left(\beta\right)R_-\left(\beta\right)}$. Then the topological invariant called winding number $w$ is defined as
\begin{equation}
w=-\frac{1}{2\pi}\frac{\left[\arg R_+\left(\beta\right)-\arg R_-\left(\beta\right)\right]_{C_\beta}}{2},
\label{eqBul2}
\end{equation}
where $\left[\arg R_\pm\left(\beta\right)\right]_{C_\beta}$ means the change of the phase of $R_\pm\left(\beta\right)$ as $\beta$ goes along $C_\beta$ in a counterclockwise way. Let $\ell_\pm$ denote the loops on the complex plane drawn by $R_\pm\left(\beta\right)$ when $\beta$ goes along $C_\beta$ in a counterclockwise way. Then the values of $w$ are determined by the number of times that $\ell_\pm$ surround the origin ${\cal O}$. We note that $w$ is not well defined when either $\ell_+$ or $\ell_-$ passes ${\cal O}$, which means that the system is gapless.

Now we demonstrate the bulk-edge correspondence for this winding number $w$ in the non-Hermitian SSH model with the matrix (\ref{eqNon10}). With the values of the parameters as $\left(t_3,\gamma_1,\gamma_2\right)=\left(1/5,5/3,1/3\right)$, we obtain the phase diagram on the $t_1$-$t_2$ plane as shown in Fig.~\ref{fig5}(a). In this phase diagram, the white region represents a normal insulator (NI) with $w=0$, and the blue region does a topological insulator (TI) phase with $w=1$. For example, at the red dot in Fig.~\ref{fig5}(a), from the trajectories $\ell_\pm$ as shown in Fig.~\ref{fig5}(b), one can find that the value of $w$ is $1$ since both $\ell_+$ and $\ell_-$ surround simultaneously the origin. Hence we expect that the topological edge states appear in the TI phase. In fact, in the energy levels in a finite open chain with these parameters along the black arrow in Fig.~\ref{fig5}(a), we can confirm the appearance of the edge states (red in Fig.~\ref{fig5}(d)) as expected. We note that the continuum energy bands in terms of the GBZ (Fig.~\ref{fig5}(c)) agree with these energy levels except for the topological edge states. In conclusion, we can establish the bulk-edge correspondence between the topological invariant defined by the GBZ and the existence of the topological edge sates in the non-Hermitian SSH model.

%
%

\section{\label{sec5}Topological semimetal phase with exceptional points}
\begin{figure}[!h]
\centering
\includegraphics[width=10cm]{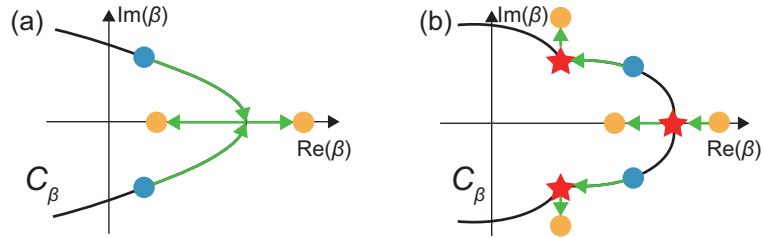}
\caption{\label{fig6}(a) Coalescence of the exceptional points. (b) Annihilation of the exceptional points at the cusp. The blue dots express the solutions of the gap-closing condition $\det{\cal H}\left(\beta\right)=0$ on the generalized Brillouin zone $C_\beta$, meaning that these are the exceptional points. On the other hand, the yellow dots express those not on $C_\beta$. The red stars are the cusps.}
\end{figure}
The phase diagram in Fig.~\ref{fig5}(a) also has a gapless phase (orange region). In fact, this gapless phase appears because of its topological stability, inherent in 1D non-Hermitian systems with the SLS and time-reversal symmetry (TRS) like the present model (\ref{eqNon7}), and so it is a topological semimetal phase (TSM). Furthermore this phase appears as an intermediate phase between the NI phase and the TI phase characterized by the winding number (\ref{eqBul2}). We note that for a real-space Hamiltonian (\ref{eqNon1}), the TRS is defined as ${\cal T}H^\ast{\cal T}^{-1}=H$, where ${\cal T}$ is a unitary matrix satisfying ${\cal T}{\cal T}^\ast=+1$.

We describe the reason why the SLS and TRS stabilize this TSM phase. Thanks to the SLS, the matrix ${\cal H}\left(\beta\right)$ can be written as an off-diagonal form (\ref{eqBul1}), and the gap of the systems closes at $E=0$. We can obtain the condition for the gap closing as $\det{\cal H}\left(\beta\right)=0$ from the characteristic equation $\det\left[{\cal H}\left(\beta\right)-E\right]=0$, and it is decomposed into two equations $\det R_\pm\left(\beta\right)=0$. These equations are polynomials of $\beta$ and $\beta^{-1}$ with real coefficients because of the TRS. Hence $\det R_\pm\left(\beta\right)=0$ can have solutions of complex-conjugate pairs, $(\beta,\beta^\ast)$. If we suppose $\beta_M$ and $\beta_{M+1}$ form a pair of the complex-conjugate solutions of $\det R_+\left(\beta\right)=0$ (or $\det R_-\left(\beta\right)=0$), they satisfy Eq.~(\ref{eqNon6}), meaning that $E=0$ is in the continuum energy bands, and the gap closes. Therefore the gap remains zero as long as this pair gives $M$th and $(M+1)$th largest absolute values among the $2M$ solutions of the equation $\det{\cal H}\left(\beta\right)=0$. In other words, as the values of system parameters changes, the GBZ is deformed so as to keep the system gapless. Thus it is unique to non-Hermitian systems.

According to the above discussion, the matrix (\ref{eqBul1}) becomes the Jordan normal form at points on the GBZ where the gap closes. This means that these points are exceptional points, where some energy eigenvalues become degenerate and the corresponding eigenstates coalesce. Importantly, we can relate the motion of the exceptional points as shown in Figs.~\ref{fig6}(a) and (b) to the change of the value of the winding number $w$. Namely, when the creation is by the inverse process of Fig.~\ref{fig6}(a) and the annihilation is by the process of Fig.~\ref{fig6}(b) (or vice versa), the systems undergo the topological phase transition from the NI phase with $w=0$ to the TI phase with $w=1$ (or vice versa). We show the detail of this discussion in Ref.~\cite{32}.

%
%

\section{\label{sec6}Summary}
In summary, we reviewed the non-Bloch band theory in 1D non-Hermitian systems. We explain how to construct the GBZ, which is given by the trajectories of $\beta_M$ and $\beta_{M+1}$ satisfying the condition $\left|\beta_M\right|=\left|\beta_{M+1}\right|$ for continuum energy bands, and show that the Bloch wave number becomes complex in an infinite open chain in general. In addition, in non-Hermitian systems, the bulk-edge correspondence between the topological invariant and existence of the topological edge states is established by using the GBZ.

We can also show that in 1D non-Hermitian systems with the SLS and TRS, the TSM phase with exceptional points appear in terms of the non-Bloch band theory, and it is regarded as an intermediate phase between the NI and TI phases. Therefore the TSM phase is stable, unlike Hermitian systems. Thus non-Hermiticity brings about qualitative changes to the topological phase transition.

Finally we mention the experimental observation of the non-Hermitian skin effect in various systems. The previous work~\cite{27} experimentally realized a nonreciprocal tight-binding model in a classical spring-mass system similarly to the simple model (\ref{eqSim1}), and observed spatially asymmetric standing waves. After that, in Ref.~\cite{28}, realizing the non-Hermitian SSH model (\ref{eqNon7}) with $\gamma_2=t_3=0$, the non-Hermitian skin effect was demonstrated by investigating the non-unitary quantum walk dynamics. Furthermore the previous work~\cite{29} experimentally also realized the non-Hermitian SSH model (\ref{eqNon7}) with $\gamma_2=t_3=0$ by using an electric circuit. It showed the differences between the energy spectra in a periodic chain and those in an open chain through an observation of the complex admittance.

%
%

\section*{Acknowledgment}
This work was supported by JSPS KAKENHI (Grants No.~JP18J22113, No.~JP18H03678, and No.~20H04633), and by the MEXT Elements Strategy Initiative to Form Core Research Center (TIES).


%

\end{document}